\documentclass{PoS}

\usepackage{dcolumn}

\newcommand{\m}{\mathcal}
\newcommand{\ord}{\alpha\alpha_s^2}
\newcommand{\Oa}{\m{O}(\ord)}
\newcommand{\ra}{\rightarrow}

\newcommand{\GeV}{\mathrm{GeV}}
\newcommand{\MeV}{\mathrm{MeV}}

\title{Electroweak Corrections at the LHC with {\tt MCFM}}

\ShortTitle{Electroweak Corrections at the LHC with {\tt MCFM}}

\author{John M. Campbell \\
        Fermilab, PO Box 500, Batavia, IL 60510, USA\\
        E-mail: \email{johnmc@fnal.gov}}

\author{Doreen Wackeroth \\
  Department of Physics, SUNY at Buffalo, Buffalo, NY 14260, USA\\
  E-mail: \email{dow@ubpheno.physics.buffalo.edu}}

\author{\speaker{Jia Zhou} \\
       Department of Physics, SUNY at Buffalo, Buffalo, NY 14260, USA\\
       E-mail: \email{jiazhou@buffalo.edu}}

     \abstract{Electroweak (EW) corrections at the LHC can be enhanced
       at high energies due to soft/collinear radiation of $W$ and $Z$
       bosons, being dominated by Sudakov-like corrections in the form
       of $\alpha_W^l\log^n(Q^2/M_W^2)\;(n\le 2l, \alpha_W =
       \frac{\alpha}{4\pi\sin\theta_W^2})$ when the energy scale $Q$
       enters the TeV regime. Thus, the inclusion of EW corrections in
       LHC predictions is important for the search of possible signals
       of new physics in tails of kinematic distributions. EW
       corrections should also be taken into account in virtue of
       their comparable size ($\m{O}(\alpha)$) to that of higher order
       QCD corrections ($\m{O}(\alpha_s^2)$).  We calculated the
       next-to-leading-order (NLO) weak corrections to the
       neutral-current (NC) Drell-Yan process, top-quark pair
       production and di-jet producion, and implemented
       them in the Monte-Carlo program {\tt MCFM}. This enables a
       combined study with the corresponding NLO QCD corrections. We
       provide both the full NLO weak corrections and their weak
       Sudakov approximation valid at high energies. The latter is
       often used for a fast evaluation of weak effects, and having
       the exact result available as well allows to quantify the
       validity of the Sudakov approximation.}

\FullConference{XXIII International Workshop on Deep-Inelastic Scattering\\
        27 April - May 1 2015\\
        Dallas, Texas}

\begin{document}

\section{Introduction}
\noindent
As the LHC reaches an unprecedented high energy and high precision,
the inclusion of electroweak (EW) corrections becomes increasingly
important in testing the Standard Model (SM) and searching for signals
of new physics, in particular in the high-energy and high-momentum
regimes of kinematic distributions. Electroweak corrections may also
play a significant role in the extraction of parton distribution
functions (PDF), for instance in constraining the gluon PDF at high
momentum fraction in di-jet production. However, the calculations of
EW corrections to relevant processes are often not readily available
in public codes and can quickly become complicated (and CPU intensive)
for high multiplicities. Even with the increasing availability of
automated tools for the calculation of EW corrections, a dedicated and
efficient computation for specific processes which are treated at the
same footing as QCD corrections in a widely used Monte Carlo program
such as {\tt MCFM}~\cite{mcfm} is still highly desirable for LHC studies.  As a
first step to improve predictions for the LHC at high energies, one
could implement the Sudakov approximation of EW corrections (see,
e.g., \cite{Butterworth:2014efa} for a review). One example of such an application is the
weak Sudakov corrections to $Z + \le 3$ jets implemented in {\tt
  ALPGEN}~\cite{alpgen}. Our goal is to implement weak corrections in
{\tt MCFM}, so that they become readily available to the experimental
community and can be studied together with the already implemented QCD
corrections. So far, we implemented the weak 1-loop corrections to the
neutral-current (NC) Drell-Yan process, $pp \to \gamma, Z \to l^+
l^-$, top-quark pair production, and di-jet production.  The
implementation of these processes in {\tt MCFM} includes both the
Sudakov approximation for weak corrections valid at high energies and
the complete 1-loop weak corrections to be able to quantify the
goodness of the Sudakov approximation. Earlier calculations and
studies of weak 1-loop corrections to these processes can be found in
Refs. \cite{nc-dy,ttb,dijet}, which we used to cross-check our
calculation and implementation. The general algorithm of
Denner-Pozzorini \cite{sudakov} is adopted in the implementation of
the weak Sudakov approximation in {\tt MCFM}. Here we present
preliminary results of the implementation of weak corrections in {\tt
  MCFM} for the processes under study for a number of relevant
kinematic distributions. In particular, in Sections \ref{sect:dy},
\ref{sect:ttb}, and \ref{sect:dijet} we compare the results of the
Sudakov approximation with the ones of the complete EW 1-loop
calculation for NC Drell-Yan, top-pair and di-jet production, respectively.

\section{Implementation of Weak Corrections in MCFM}
\begin{figure}
   \vspace{-5mm}
   \begin{center}
     \hspace{-5cm}
     \includegraphics[scale=0.34]{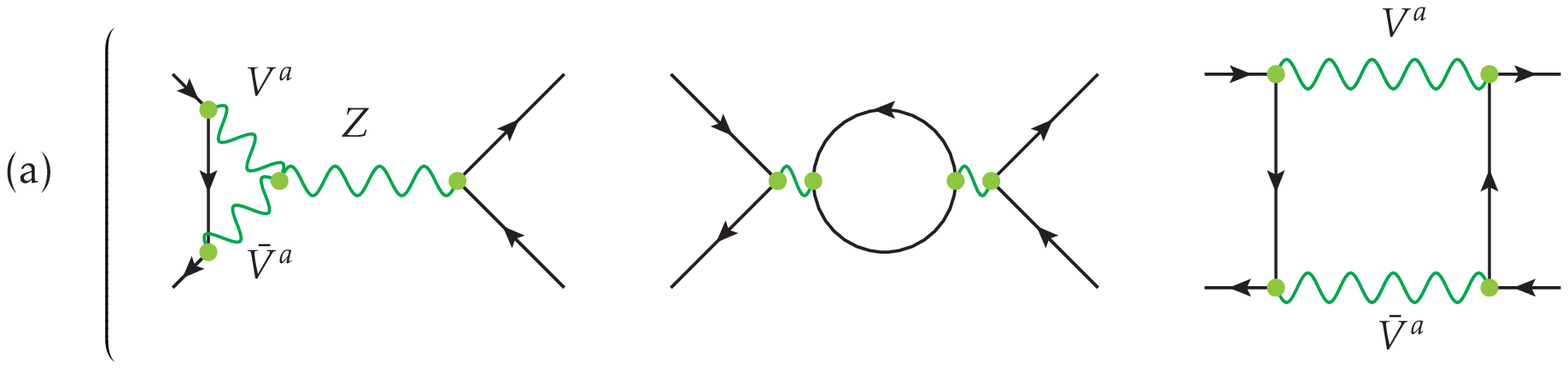}
   \end{center}
   \vspace{-8.6cm}
   \begin{center}
     \hspace{-8mm}
     \includegraphics[scale=0.34]{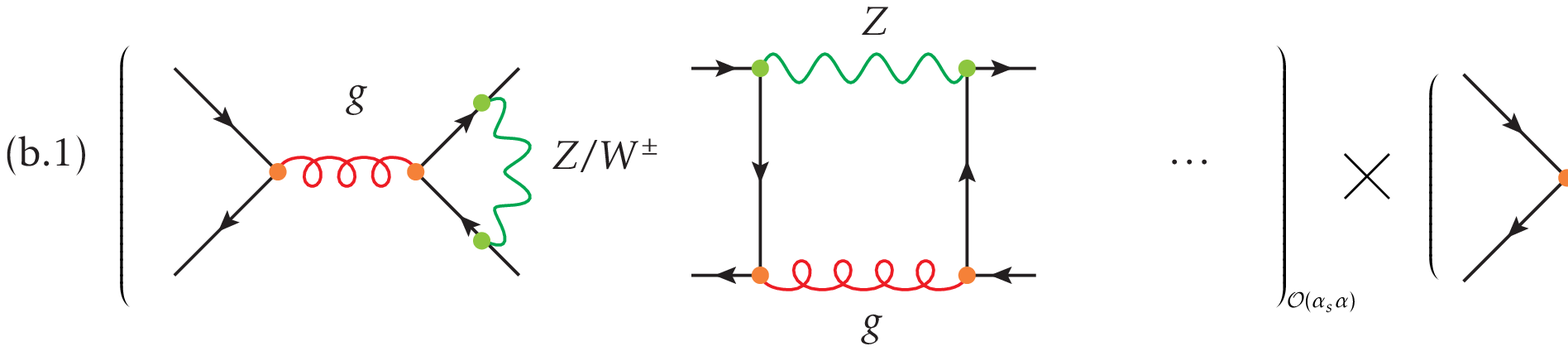}
     \hspace{8mm}
   \includegraphics[scale=0.34]{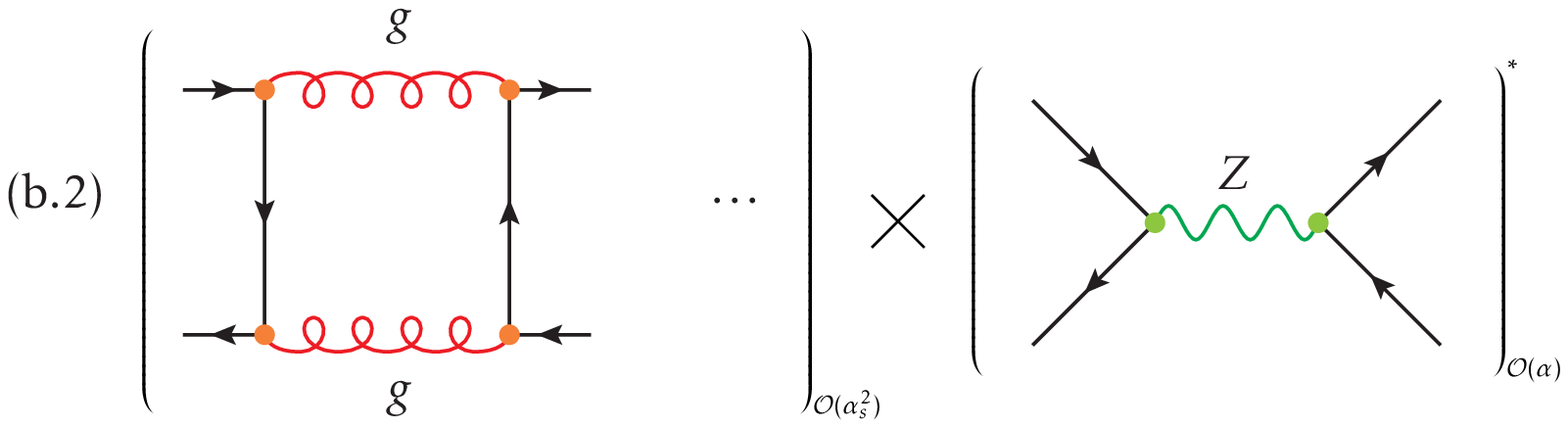}
   \end{center}
   \vspace{-8.8cm}
   \begin{center}
     \hspace{-1.5cm}
     \includegraphics[scale=0.4]{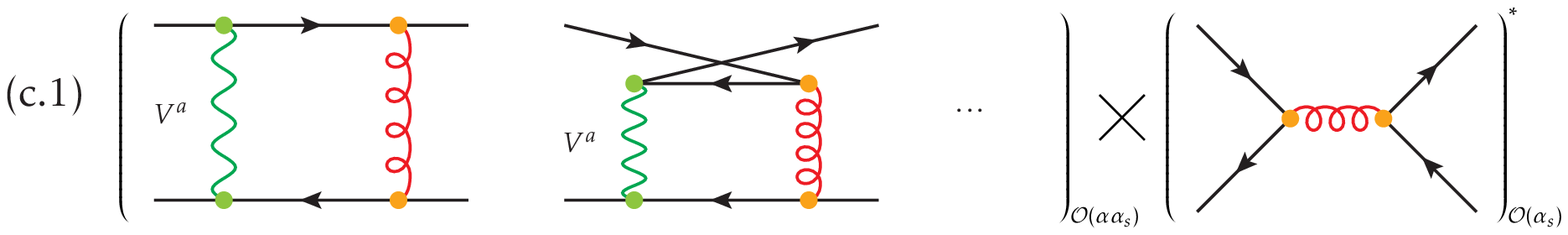}
     \hspace{-1cm}
     \includegraphics[scale=0.4]{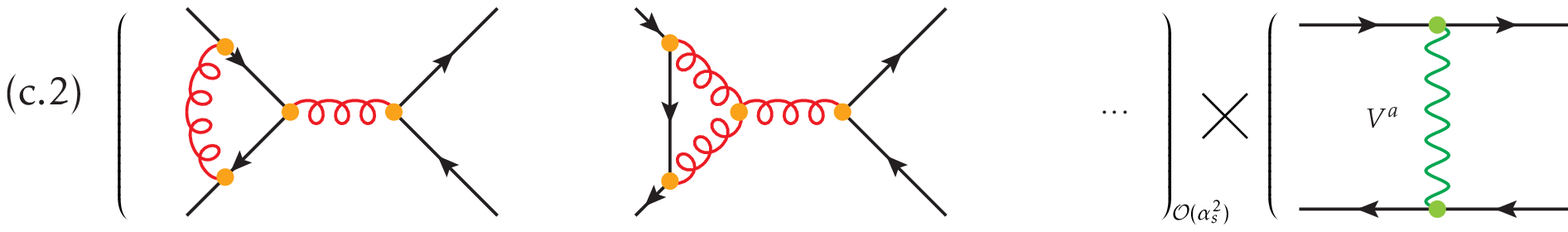}
   \end{center}
   \vspace{-9cm}
   \caption{\label{fig:nlo:sample} Sample Feynman diagrams for the
     $\m{O}(\alpha)$ weak corrections to the LO cross sections of
     the NC Drell-Yan process (a), top-quark pair production (b.1),
     and di-jet production (c.1). Figures (b.2) and (c.2)
     illustrate $\m{O}(\alpha_s)$ QCD corrections to the interference
     of QCD-EW mixed LO cross sections which arise in the case of four-quark
     external states in top-quark pair and di-jet production.}
 \end{figure}

\noindent
We investigate the SM weak 1-loop corrections to the processes
discussed in Sect. \ref{sect:dy}, \ref{sect:ttb}, and
\ref{sect:dijet}, i.e., the $\m{O}(\alpha)$ corrections that include
the virtual contribution of massive gauge $V^a\,(Z/W^\pm)$ and Higgs
bosons in higher order diagrams.  Figure \ref{fig:nlo:sample} shows
sample Feynman diagrams that contribute to these relative (virtual)
corrections in NC Drell-Yan, top-quark pair and di-jet production. In
the latter two processes, however, we need to take into account also
NLO QCD corrections to the interference of the QCD-EW mixed leading-order (LO)
contribution to achieve the same order of $\Oa$, which are illustrated
in Fig.\ref{fig:nlo:sample} (b.2) and (c.2). These NLO QCD virtual
corrections require the inclusion of the real correction with
emission/absorption of a gluon to cancel the IR divergence. On the
other hand, photonic corrections are not included since they
themselves form a separate gauge invariant subset and are not enhanced
at high-energy scales.

\noindent 
We use the on-shell renormalization scheme and a constant gauge boson decay width
in the $Z/W^\pm$ propagators in the calculation. 
To produce the numerical results we use the following input parameters:
\begin{eqnarray}
  && G_{\mu} = 1.16639~\times~10^{-5}\,\GeV^{-2}, \;
  \alpha_{\mu} = 1/132.5605045, \;
  \sin^2\theta_W = 1 - M_W^2/M_Z^2, \; \nonumber \\
  &&M_Z = 91.1876\,\GeV, \;  
  \Gamma_Z = 2.4952\,\GeV, \;
  M_W = 84.425\,\GeV, \;
  \Gamma_W = 2.141\,\GeV, \; \nonumber \\
  &&M_H = 120\,\GeV, \;
  m_t = 173.2\,\GeV, \;
  m_b = 4.6\,\GeV, \;
  m_u = 66\,\MeV, \;
  m_d = 66\,\MeV, \nonumber \\
  &&m_c = 1.2\,\GeV, \;
  m_s = 150\,\MeV, \;
  m_e = 0.51099892\,\MeV, \;
  m_{\mu} = 105.658369\,\MeV, \nonumber \\
  &&m_{\tau} = 1.777\,\GeV. \nonumber
\end{eqnarray}
\noindent
Note that the quarks and leptons
except for the top quark are treated massless if those particles are
initial/final-state particles. We only retain their masses in closed fermion
loops, i.e. when calculating the gauge boson self-energy corrections in the NC
Drell-Yan process.
The factorization scale $\mu_F$ is set to be equal to the
renormalization scale $\mu_R$, and is chosen to be the mass of the $Z$
boson $M_Z$, twice of the mass of the top quark $2 m_t$, and the
transverse momentum of the leading jet $p_{T,j_1}$, in NC Drell-Yan,
top-pair and di-jet production, respectively.

\section{Neutral Current Drell-Yan Process}\label{sect:dy}
\noindent
The partonic level process under consideration is $q\bar{q} \ra
l\bar{l}$ with exchange of a neutral EW gauge boson ($\gamma/Z$),
where $q \in \{u,d,c,s,b\}$ denotes initial-state quarks, and $l \in
\{e,\mu,\tau;\nu_{e},\nu_{\mu},\nu_{\tau}\}$ final-state leptons. The
LO cross section is thus of $\m{O}(\alpha^2)$. When we
consider NLO weak corrections to the LO
process, it refers to a weak gauge boson exchange in vertex,
self-energy, and box corrections as demonstrated in
Fig.\ref{fig:nlo:sample} (a). In order to better understand and
characterize the validity of the Sudakov approximation, we produce
differential distributions of relative corrections with respect to the
invariant mass $M_{l^+l^-}$ and transverse momentum $p_{T,l^+(l^-)}$
using the exact 1-loop calculation and Sudakov approximation.  The
relative correction $\delta$ is defined as follows,
\begin{equation}{\label{relcorr}}
    \delta = \frac{\delta\sigma^{1-loop}}{\sigma_{LO}}
    = \frac{\sigma_{NLO} - \sigma_{LO}}{\sigma_{LO}},
\end{equation}
where $\sigma_{LO}$ and $\sigma_{NLO}$ are the LO and NLO cross sections. 

\begin{figure}
  \begin{center}
\includegraphics[scale = 0.26,angle = -90]{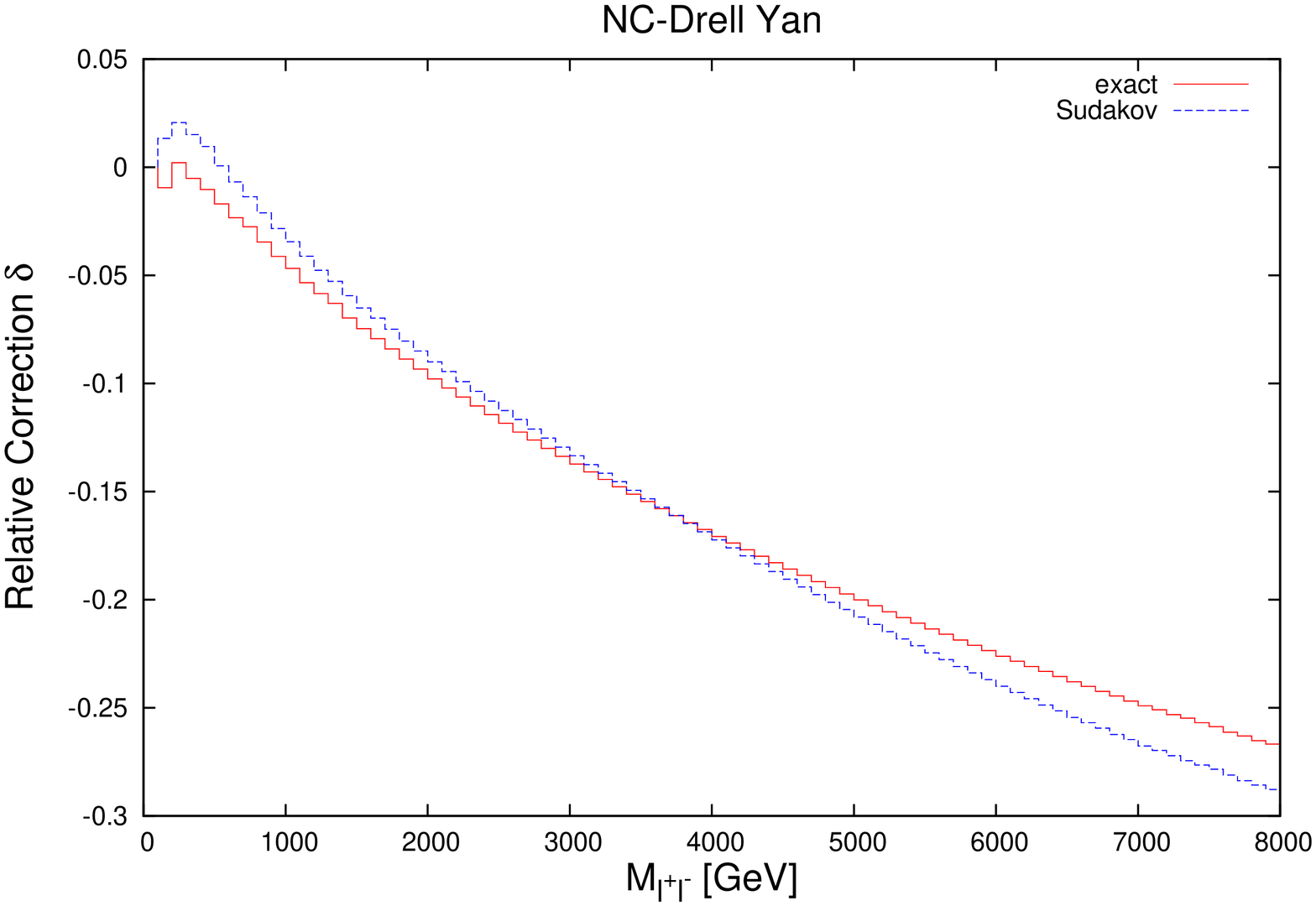}
\includegraphics[scale = 0.26,angle = -90]{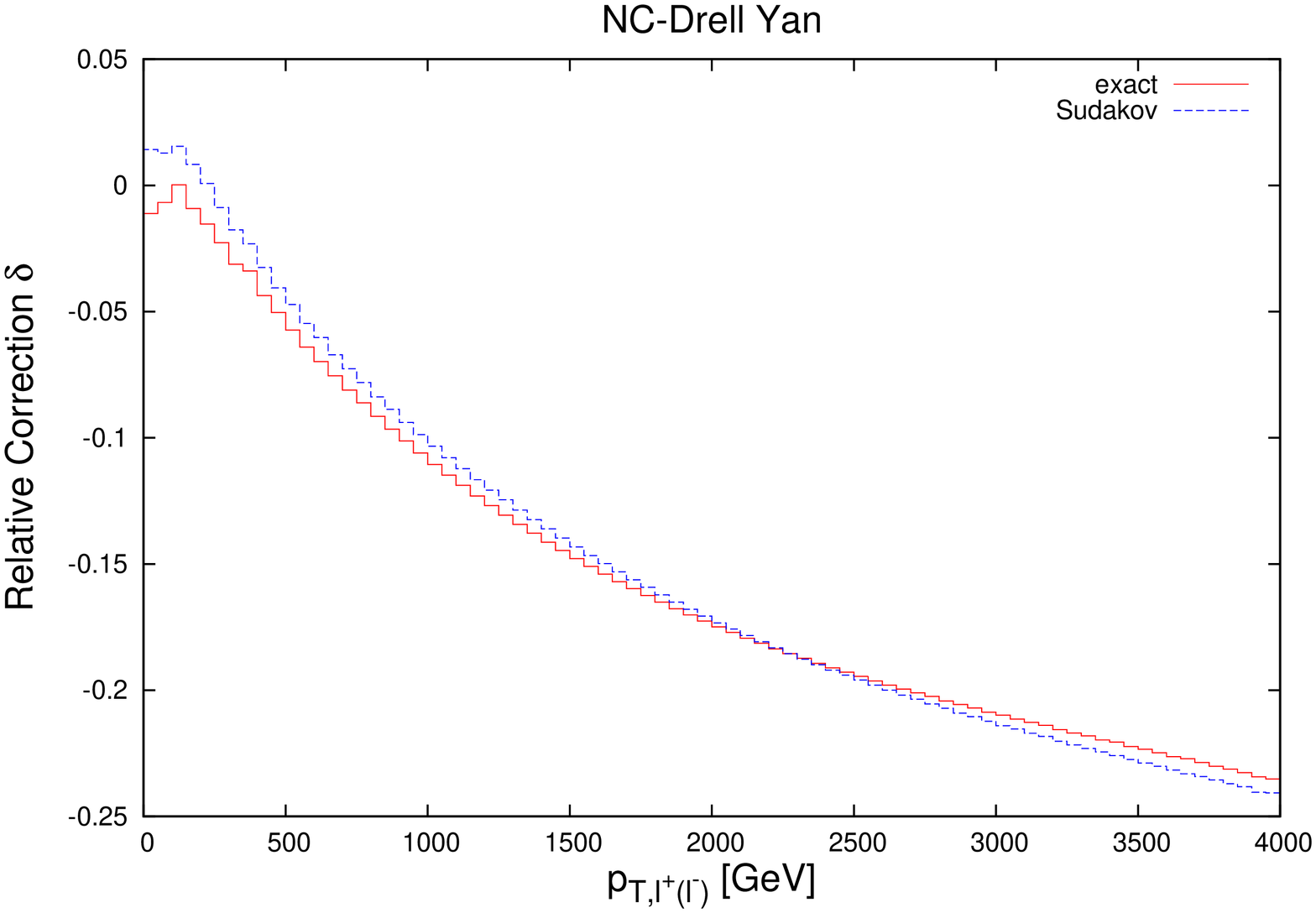}
\end{center}
\caption{\label{fig:dy} The relative correction $\delta$ to the
  invariant mass $M_{l^+l^-}$ (left) and the transverse momentum
  $p_{T,l^+(l^-)}$ (right) distributions at the LHC at $\sqrt{s}=14$
  TeV, for the NC Drell-Yan process $pp\ra~l^+l^-$. In
  each plot, the red curve denotes the distribution obtained with the
  exact NLO calculation, while the blue one denotes the one obtained
  with the Sudakov approximation. No cuts have been applied.}
\end{figure}

\noindent
Figure~\ref{fig:dy} shows $\delta$ for both distributions for $pp \to
\gamma,Z \to l^+ l^-$ at the LHC at $\sqrt{s}=14$ TeV. The results are
produced with the PDF set MRST2004QED \cite{mrstqed}.  As can be seen
the Sudakov approximation shows good agreement with the exact NLO
calculation in both distributions, and the largest discrepancy which
appears in the invariant mass distribution is as small as a few
percent ($\sim 2\% - 3\%$) at 8 TeV.

\section{Top-Quark Pair Production}\label{sect:ttb}
\begin{figure}
\begin{center}
\includegraphics[scale = 0.26, angle = -90]{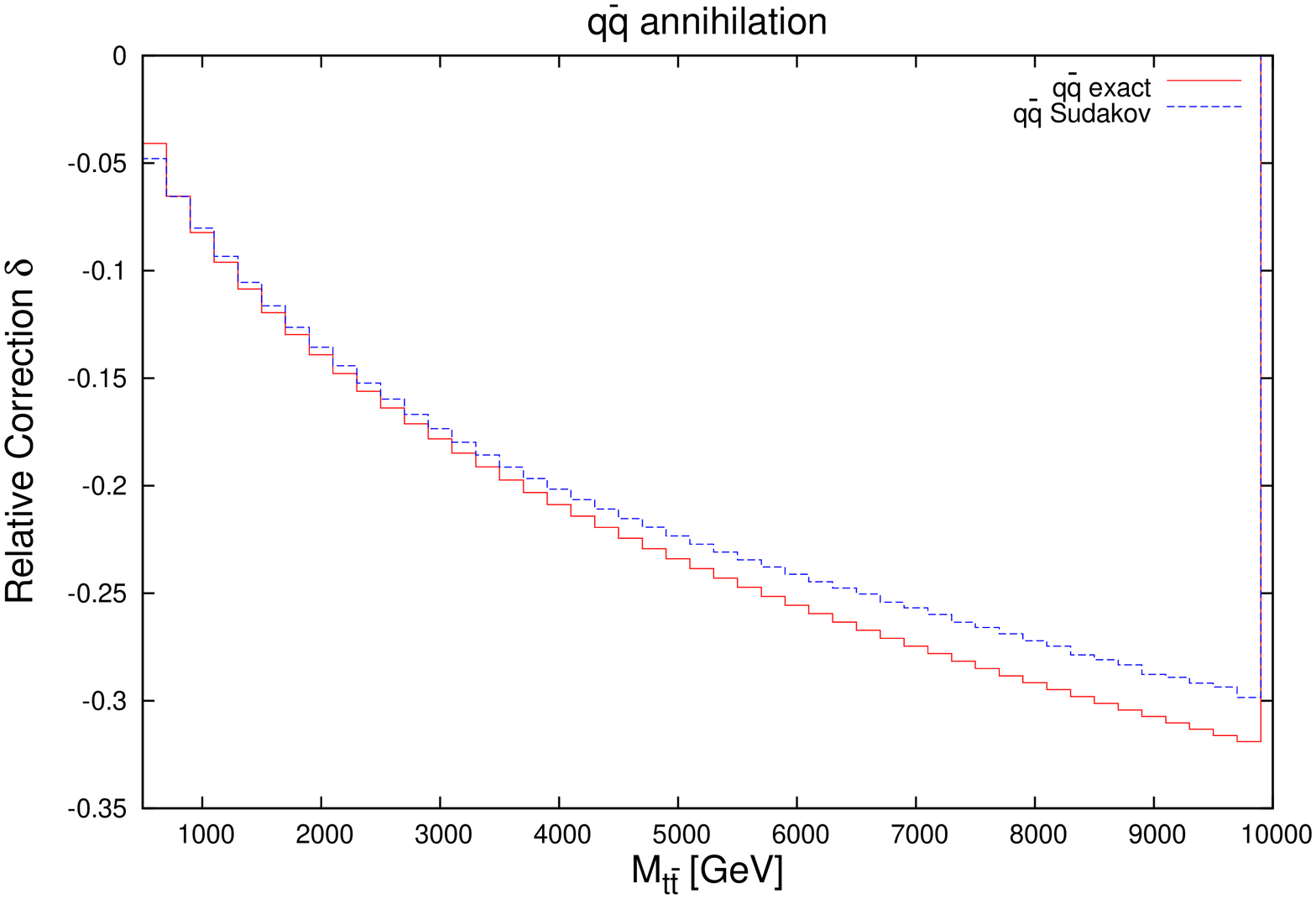}
\includegraphics[scale = 0.26, angle = -90]{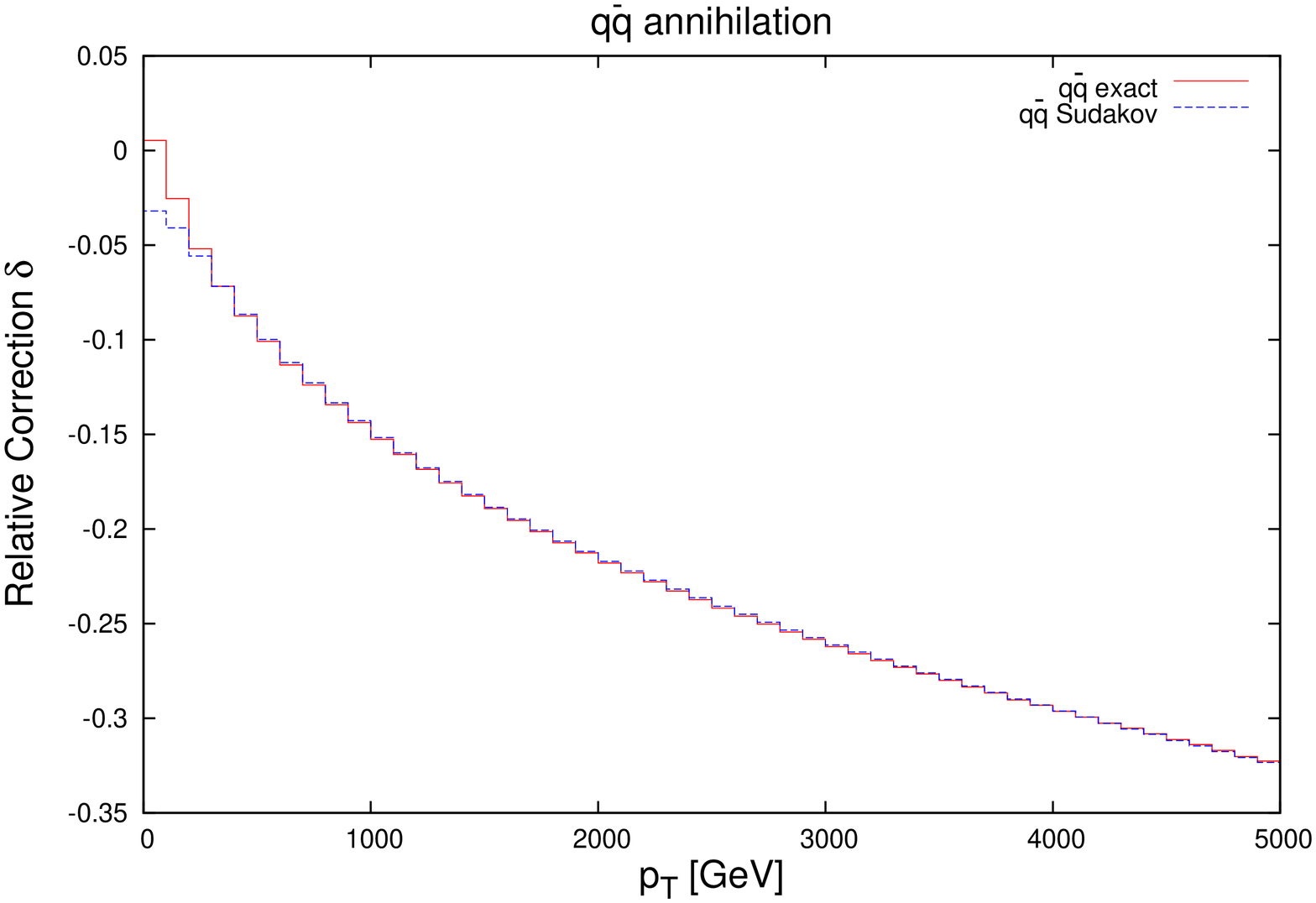}
\includegraphics[scale = 0.26, angle = -90]{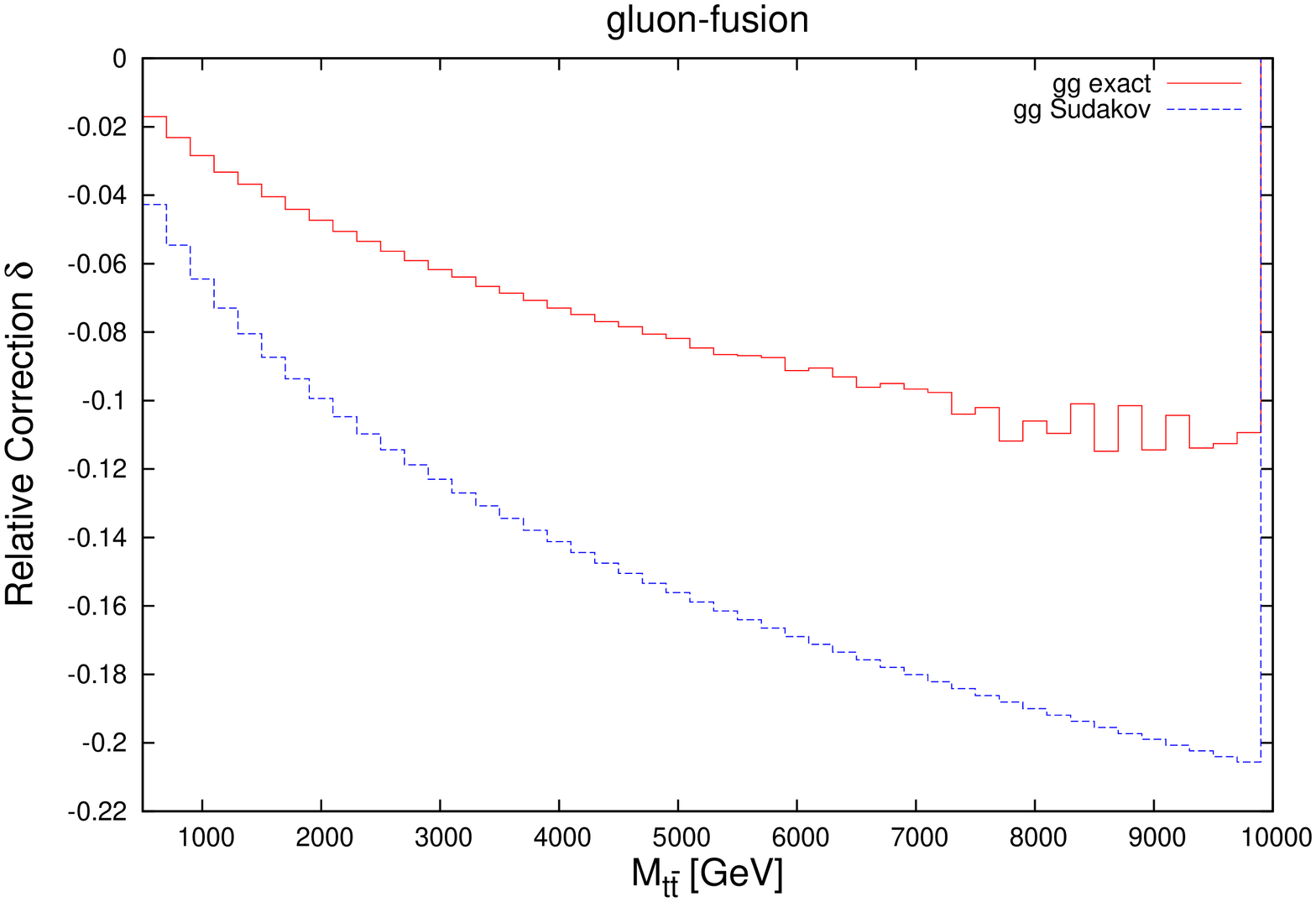}
\includegraphics[scale = 0.26, angle = -90]{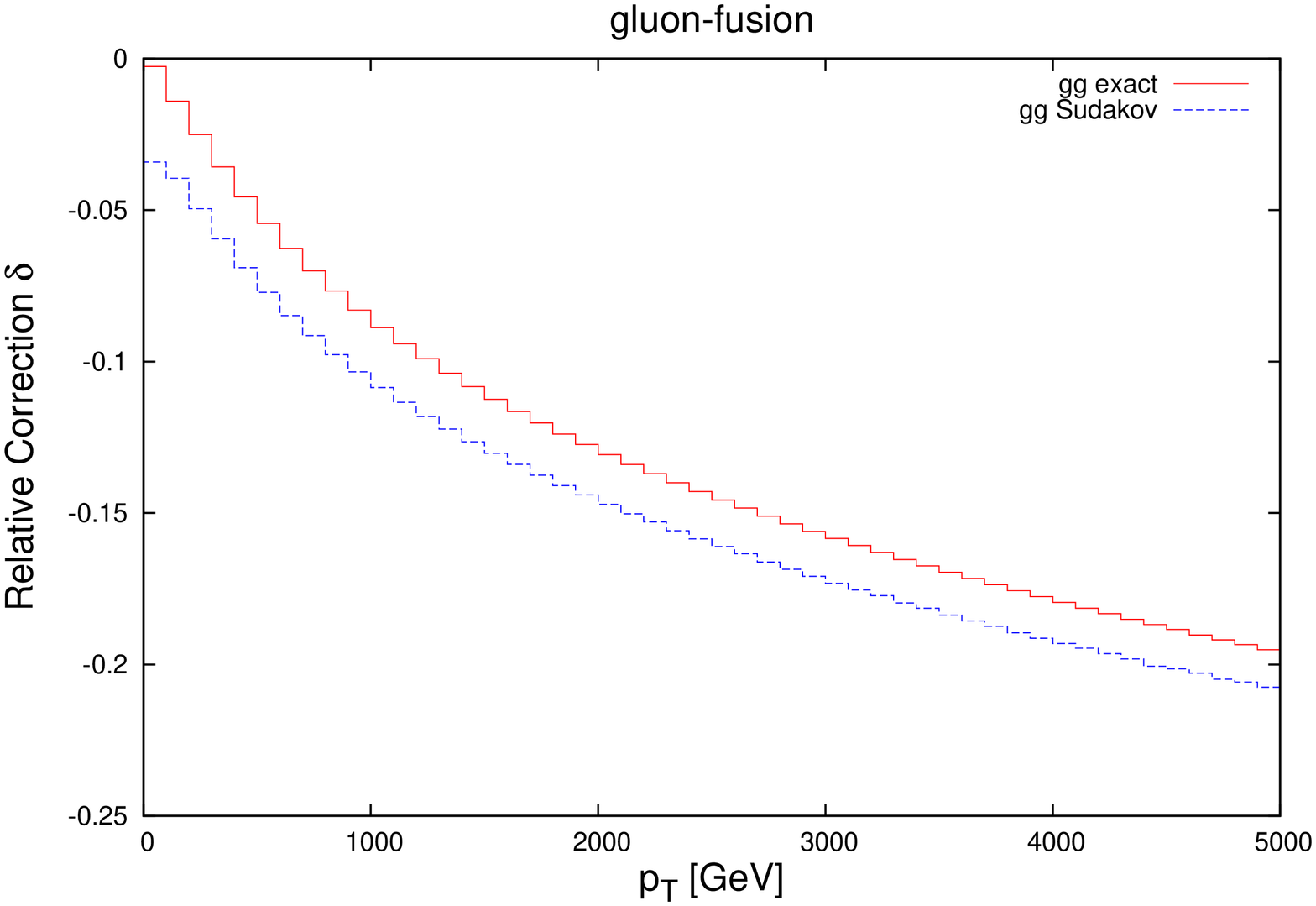}
\end{center}
\caption{\label{fig:ttb} The relative correction $\delta$ to the
  invariant mass $M_{t\bar{t}}$ and transverse momentum
  $p_{T,t(\bar{t})}$ distributions in top-quark pair production at the
  LHC at $\sqrt{s} = 14$ TeV. No cuts have been applied. The upper and lower two
  plots show respectively the results for the quark-antiquark annihilation and gluon fusion channel.
Again the red curves denote the exact NLO calculation and the blue curves the
  Sudakov approximation.}
\end{figure}

\begin{figure}
  \centering
  \includegraphics[scale = 0.3, angle = -90]{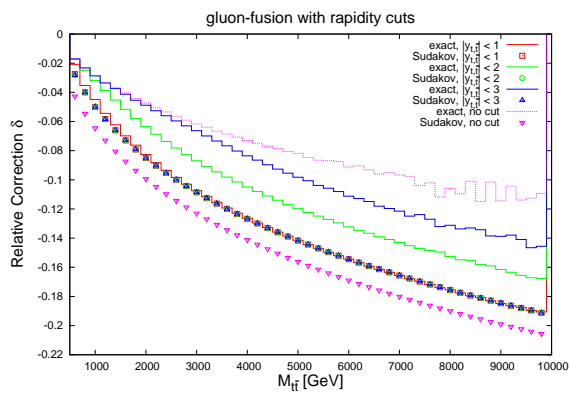}
  \caption{\label{fig:ttb:ggcut} The relative correction $\delta$ to the 
    $M_{t\bar{t}}$ distribution in top-quark pair production at
    the LHC at $\sqrt{s} = 14$ TeV for different cuts on $|y_{t,\bar{t}}|$. The results represented by
    symbols are those of the Sudakov approximation and the lines indicate the results
    obtained with the exact NLO calculation.}
\end{figure}

\noindent
At the LHC, top-quark pairs are dominantly produced via strong
quark-antiquark annihilation and gluon fusion at the parton level,
with an LO cross section of $\m{O}(\alpha_s^2)$. We consider the
1-loop weak contribution of $\Oa$ to these
top-quark pair production processes. In addition to the 
weak 1-loop corrections that are present in both channels,
the quark-antiquark annihilation channel includes
the QCD corrections to the interference of QCD-weak mixed LO
contributions.  These are comprised of the virtual box corrections shown in
Fig.\ref{fig:nlo:sample} (b.2), as well as the corresponding real corrections
with emission/absorption of a gluon. We use the PDF set CTEQ6L \cite{cteq6}
to produce the numerical results in Fig.\ref{fig:ttb} and
Fig. \ref{fig:ttb:ggcut}.

\noindent
Figure~\ref{fig:ttb} shows the results for the relative corrections at
the LHC at $\sqrt{s} = 14$ TeV, where the upper(lower) two plots show
the results for the invariant mass $M_{t\bar{t}}$ and transverse
momentum $p_{T,t(\bar{t})}$ distributions in the quark-antiquark
annihilation(gluon fusion) channel.  As expected the Sudakov
approximation works better in case of quark-antiquark annihilation,
since this subprocess is analogous to the NC Drell-Yan at high
energies where the effect of the mass of the top quark is
negligible. In contrast, in the gluon fusion channel, there is an
obvious discrepancy between the Sudakov approximation and the exact
NLO calculation in the $t\bar t$ invariant mass distribution while in the
top-quark transverse momentum distribution, it is not as significant. After
further investigation, it is clear that this disagreement is caused
by the mismatch of the angular dependence because there is no such
information captured by the Sudakov
approximation in the gluon fusion channel. Since the transverse
momentum distribution has less dependence on the scattering angle, or
equivalently rapidity, it is expected that the Sudakov approximations works better in this case. 
The invariant $t\bar t$ mass is a function of rapidity, 
$M_{t\bar{t}}^2 = 2~m_t^2 + 2~m_T^2\cosh(y_t - y_{\bar{t}}) + 2~p_T^2$,
where $m_T = \sqrt{p_T^2+m_t^2}$, and $y_{t(\bar{t})}$ denotes
rapidity. We therefore expect to find
a better agreement between the Sudakov approximation and the exact result by 
imposing a cut in the top-quark rapidity. This is illustrated in Fig.\ref{fig:ttb:ggcut} where we show results
for different rapidity cuts. We indeed find agreement when $|y_{t,\bar{t}}| <
1$. This constrains the range of validity of the Sudakov approximation,  as has been also
pointed out in Ref.~\cite{ttb2}.

\section{Di-jet Production}\label{sect:dijet}
\noindent
Di-jet production is described at LO by $2 \to 2$ processes involving quarks and gluons,
where the two final-state partons are detected as jets after hadronization. We thus
categorize the processes contributing to di-jet production into three subprocesses according to
the number of external quarks or gluons, which are four-quark,
two-gluon-two-quark, and four-gluon subprocesses. At tree
level, these processes can be produced via strong or EW
interactions. Thus, the LO cross section in di-jet
production consists of the purely QCD contributions of
$\m{O}(\alpha_s^2)$, the mixed QCD-EW contribution of
$\m{O}(\alpha_s\alpha)$, as well as the purely EW contribution of
$\m{O}(\alpha^2)$. The NLO corrections consist of virtual and real
corrections, where the latter exist in the four-quark subprocesses
associated with a emitted/absorbed gluon. And there are two types of
the interference of diagrams that contribute to the full NLO
corrections of $\Oa$, and symbolically, we can write them as 
\begin{eqnarray}\label{sig_nlo:propto}
\hat{\sigma}(\ord) \propto 
\left\{
\begin{array}{l}
2\mathrm{Re}\left[\m{M}(\alpha_s\alpha)\cdot\m{M}^*(\alpha_s)\right] \\
2\mathrm{Re}\left[\m{M}(\alpha_s^2)\cdot\m{M}^*(\alpha)\right]
\end{array}
\right.
\end{eqnarray}
where $\m{M}(\alpha_s)$ and $\m{M}(\alpha)$ denote the LO
amplitude with gluon and weak boson exchange, respectively, and
$\m{M}(\alpha_s^2)$ the NLO QCD correction to the strong LO
amplitude. In the virtual corrections, $\m{M}(\alpha_s\alpha)$ denotes
the NLO amplitude that can be either weak correction to the strong LO
amplitude, or QCD correction to the weak LO amplitude; while in the
real correction, the NLO amplitude $\m{M}(\alpha_s\alpha)$ is
restricted to the latter because we only need gluon radiation in the
real contribution to cancel the IR divergence in the
virtual contribution. Note that there is no NLO weak correction to the four-gluon
subprocesses.

\begin{figure}
  \centering
  \includegraphics[scale = 0.3, angle=-90]{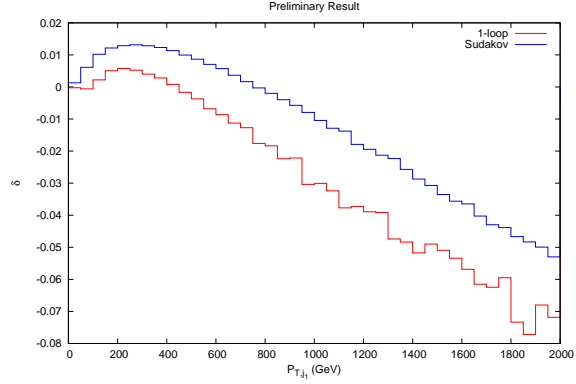}
  \caption{\label{fig:dijet} The relative
    correction $\delta$ to the transverse momentum distribution of the leading jet
    $p_{T,j_1}$ in di-jet production at the LHC at $\sqrt{s} = 14$ TeV. Again the
  red curve denotes the exact NLO calculation and the blue curve the
  Sudakov approximation.}
\end{figure}

\noindent
We calculate the inclusive jet process using the anti-$k_T$ jet clustering
algorithm \cite{ankt}, and set the pseudo-cone size to $R = 0.6$. We apply the following
cuts on the jet transverse momentum and jet rapidity:
\begin{equation}\label{jetcut}
  p_{T,j} > 25\; \GeV, \;\;\; |y_j| < 2.5.
\end{equation}
We use the PDF set CTEQ6L1 \cite{cteq6}. Figure \ref{fig:dijet} shows
preliminary results for the relative correction at the LHC at $\sqrt{s} = 14$ TeV. The comparison
between the Sudakov approximation and exact NLO
correction is restricted to the one-loop weak relative correction of
$\m{O}(\alpha)$, i.e., the LO cross section in
Eq.(\ref{relcorr}) is purely QCD, and the total NLO result
does not contain the LO EW effect. We find that Sudakov
approximation gives less negative contribution than the complete set
of NLO corrections that amount to the contribution of
$\Oa$. There might be two reasons for this
disagreement between the approximation and the exact calculation. One is
the missing angular dependence in the Sudakov approximation in the
two-gluon-two-quark subprocesses, which is similar to what we observed in the
gluon fusion channel in top-quark pair production. The second is
that we neglected the QCD corrections to the EW LO amplitude in
the Sudakov approximation. Therefore, we may only be able to use the Sudakov
approximation in some particular subprocesses such as four-quark processes, 
and a cut on the scattering angle (or rapidity)
should be imposed when the two-gluon-two-quark subprocesses are
included. Final results and a more detailed discussion will be 
made available in Ref.~\cite{paper}.

\acknowledgments
This research is supported in part
by the US DOE under contract DE-AC02-07CH11359 and the NSF under award no. PHY-1118138.

\end{document}